\documentstyle[12pt]{article}
\begin{document} 
\centerline{{\large {\bf Algebraic Structure of Discrete Zero
Curvature Equations}}} 
\centerline {{\large {\bf and Master Symmetries of
Discrete Evolution Equations}}} 

\vskip 2mm \centerline {Wen-Xiu
Ma$^{\dagger}$\footnote{Email:  mawx@cityu.edu.hk} and Benno
Fuchssteiner$^\ddagger$\footnote{ Email:  benno@uni-paderborn.de}}
\centerline{$^\dagger$Department of Mathematics, City University of Hong Kong}
\centerline{Kowloon, Hong Kong, China} 
\centerline{$^\ddagger$Department of Mathematics, 
University of Paderborn, D-33098
Paderborn, Germany}

\renewcommand{\thesection}{\Roman{section}} 

\newtheorem{thm}{Theorem}
\newtheorem{cor}{Corollary}
\newtheorem{ex}{Example}
\newcommand{\R}{\mbox{\rm I \hspace{-0.9em} R}}        
\newcommand{\Z}{\mbox{\rm Z }}        
\def\be{\begin{equation}}
\def\ee{\end{equation}}
\def\ba{\begin{array}}
\def\ea{\end{array}}
\def\bea{\begin{eqnarray}}
\def\eea{\end{eqnarray}}
\def\la {\lambda}
\def \part {\partial}
\def \al {\alpha}
\def \de {\delta}

\setlength{\baselineskip}{17pt}

\begin{abstract}
An algebraic structure related to discrete zero curvature equations 
is established. It is used to give an approach for generating  
master symmetries of first degree for systems of discrete evolution equations 
and an answer to why there exist such master symmetries. 
The key of the theory 
is to generate nonisospectral flows $(\lambda _t=\lambda ^l,\ l\ge0)$
from the discrete
spectral problem associated with a given system of discrete evolution 
equations. Three examples are given.
\end{abstract}

\section{Introduction}

The theory of integrable systems has various aspects, although
the term ``integrable" is somewhat 
ambiguous, especially for systems of partial differential
equations. Symmetries are one of those important 
aspects and have a deep mathematical and physical background. 
When any special character, for example the Lax pair, hasn't been found   
for a given system of continuous or discrete equations,
among the most efficient ways is to consider its symmetries in order 
to obtain exact solutions.
It is through symmetries that Russian scientists 
et al. developed some theories for testing
the integrability of systems of evolution equations,  
both continuous and discrete,
and classified many types of systems of nonlinear equations
which possess higher differential or 
differential-difference degree symmetries (for example,
see \cite{MikhailovSS} \cite{LeviY}).
Usually an integrable system of equations
is referred as to a system possessing infinitely many symmetries 
\cite{FuchssteinerF} \cite{Fokas}. 
Moreover these symmetries form nice and interesting algebraic structures
\cite{FuchssteinerF} \cite{Fokas}.        

For a given system of evolution equations 
$u_t=K(u)$, both continuous and discrete,
a vector field $\sigma (u)$ is called its symmetry if
$\sigma (u)$ satisfies its linearized system
\be \frac {d \sigma (u)}{dt}=K'[\sigma ],\ {\rm i.e.,} \ \frac {\part \sigma}
{\part t}=[K,\sigma ]:=K'[\sigma ]-\sigma '[K],\label{sd}\ee
where the prime means the Gateaux derivative.
Starting from a Lie-point symmetry, we can often construct
the corresponding explicit group-invariant solutions.
A symmetry $\sigma$ may, 
of course, depend explicitly on the evolution variable $t$. 
If a symmetry $\sigma $ of the system $u_t=K(u)$ not depending  
explicitly on $t$ is a polynomial in $t$, i.e.,  
\be \sigma (t,u)=\sum_{i=0}^{n}\frac {t^i}{i!}\rho _i(u),\ n\ge 1,
\label{ts} \ee
then we have 
\be \rho _i=[K, \rho_{i-1}], \ 1\le i\le n,\ee
and 
\be ({\rm ad}_K)^{n+1}\rho _0=0,\ {\rm where}\ ({\rm ad}_K)\rho _0
=[K,\rho _0].\label{ms}\ee
Therefore the symmetry (\ref{ts}) is 
totally determined by a vector field $\rho _0$ satisfying
(\ref{ms}). This kind of vector field $\rho _0$
has been discussed in considerable
detail and is called a master symmetry of 
degree $n$ of $u_t=K(u)$ by one of the authors
(BF) in Ref. \cite{Fuchssteiner1983}. 

The appearance of first degree master symmetries 
gives a common character for integrable systems of 
continuous evolution equations 
both in $1+1$ dimensions and in $2+1$ 
dimensions, for example, the KdV equation and the KP equation. 
The resulting symmetries are sometimes
called $\tau$-symmetries (for more
information, see \cite{Ma1990} for example) and usually 
constitute centerless Virasoro algebras together
with time-independent symmetries 
\cite{ChenL} \cite{OrlovS-LMP1986} 
\cite{LiZ-JPA1986}. Moreover these 
$\tau$-symmetries may be generated by use of zero curvature equations 
or Lax equations \cite{Ma1993c} and the corresponding master symmetry
flows may also be solved by the inverse scattering method
\cite{CalogeroD-LNC1978} \cite{ChanZ-LMP1987}.  
In the case of systems of discrete evolution equations, there exist
some similar results. For example, many systems of discrete evolution 
equations have $\tau$-symmetries and centerless Virasoro symmetry algebras
\cite{OevelZFR} \cite{OevelFZ} \cite{ZhangTOF},
and the inverse scattering method may still be applied in solving 
themselves and their master symmetry flows 
\cite{AblowitzL-JMP19756}
\cite{DeviBR-NCA1980} 
\cite{BruschiLR-JMP1981} \cite{BruschiR-LNC1981}.         
So far, however, to the best of our knowledge,
there hasn't been a systematic mathematical theory to explain 
why there exist $\tau$-symmetries for systems of 
discrete evolution equations
and how we can construct those $\tau$-symmetries when they exist,
from the point of discrete zero curvature equations.

Throughout this paper, 
``master symmetries" is used to express the first degree master symmetries
which generate $\tau$-symmetries.
Our purpose   
is to give an algebraic explanation of the first question above
and to provide a procedure to generate those master symmetries 
for a given lattice hierarchy. The discrete zero
 curvature equation is our basic tool to give rise to our answer and 
 procedure.  
The Volterra lattice hierarchy, the Toda lattice hierarchy 
and a sub-KP lattice hierarchy are chosen and
analyzed as some illustrative examples, which have one 
dependent variable, two dependent variables and three dependent variables,
respectively.

Let us now describe our notation. 
Assume that $u=(u_1,\cdots,u_q)^T$ where $u_i=u_i(t,n),\ 1\le i\le q,$
are real functions defined over $\R \times \Z $ 
 (in the case of the complex function, the discussion
is similar), and
let $\cal B$ denote all real functions $P[u]=P(t,n,u)$ which 
are $C^\infty$-differentiable with respect to $t$ and $n$, and 
$C^\infty$-Gateaux differentiable with respect to $u$. 
We always write $E$ as a shift operator and  
\be
(E^mx)(n)=x^{(m)}(n)=x(m+n),\ \textrm{where} \ x:\Z \to \R,\  m,n\in \Z. \ee 
Note that $x^{(m)}$ here doesn't mean the $m$-th derivative. Set 
${\cal B}^r=\{(P_1,\cdots,P_r)^T|\,P_i\in {\cal B},\ 1\le i\le r\}$, 
and denote by ${\cal V}^r$ all matrix operators $\Phi=(\Phi_{ij})_{r\times r}$
where the entries $\Phi _{ij}=\Phi _{ij}(t,n,u)\in {\cal B},$ 
and by ${\widetilde  {\cal V}}^r$, 
all matrix operators depending on a parameter $\la $:
$U =(U_{ij})_{r\times r}$, 
where the entries $U  _{ij}=U_{ij}(t,n,u,\la )\in {\cal B}$ for all $\la $, 
being $C^\infty$-differentiable with respect to $\la $.

We will need a multiplication operator 
 \be [n]: {\cal B} \to {\cal B},\
 P[u]\mapsto [n]P[u],\ ([n]P[u])(m)=m(P[u])(m),\label{multio}\ee
 which is often involved in the construction of master symmetries.
 This avoids an unclear expression $nP[u]$, which may also mean
  $(nP[u])(m)=n(P[u])(m)$. For example, it is absolutely clear that
  $([n]P[u])(m)=mu(m-1)+mu(m)$, when $P[u]=E^{-1}u+u$.
We also need
a difference operator $\Delta =E-E^{-1}$, whose inverse operator
may be defined by 
\be (\Delta ^{-1}u)(n)=((E-E^{-1})^{-1}u )(n)
:=\frac 12 (\sum_{k=-\infty}^{-1}u(n+1+2k)-
\sum_{k=1}^{\infty}u(n-1+2k)),
\label{Deltainverse1}   \ee 
where $u$ is required to be rapidly vanishing at the infinity. 
Moreover we define 
\be (\Delta ^{-1}\alpha )=(1/2)\alpha [n], \ 
\textrm{i.e.,}\ (\Delta ^{-1}\alpha )(n)=(1/2)\alpha n, \ 
\alpha =\textrm{const.}
\label{Deltainverse2} \ee
Obviously we can find that
\be (E-1)^{-1}=\Delta ^{-1}(1+E^{-1}),\  
(1-E^{-1})^{-1}=\Delta ^{-1}(E+1), \label{inversed}\ee 
and thus 
\be (E-1)^{-1}\alpha =\alpha [n],\ (1-E^{-1})^{-1} \alpha=\alpha [n], \ 
\alpha =\textrm{const.},\ee 
which may also be viewed as a definition of two inverse operators 
$(E-1)^{-1}$ and $(1-E^{-1})^{-1}$.
Note that here we have used the operator $[n]$ so that 
two functions $(E-1)^{-1}\alpha $ and $(1-E^{-1})^{-1}\alpha $
have the other clear expressions.
The operators $\Delta ^{-1}$, 
$(E-1)^{-1}$ and $(1-E^{-1})^{-1}$
often appears in the expressions of master
symmetries and thus master symmetries are usually nonlocal vector fields
belonging to ${\cal B}^q$.

In order to carefully analyze algebraic structures
related to symmetries, we specify the definition of
the Gateaux derivative $X'[S]$ of 
any vector valued function $X\in {\cal B}^r$ at a
direction $S\in {\cal B}^q$ as follows
\be X'[S]=\left.\frac d {d\varepsilon}\right|_{\varepsilon =0}
X(u+\varepsilon S),\ee 
which implies that $X'$ is an operator from ${\cal B}^q$ to ${\cal B}^r$,
and need the following two product operations
\bea &&
 [K,S]=K'[S]-S'[K],\ K,S\in {\cal B}^q,\label{prodd1}\\ &&
  \lbrack\!\lbrack f, g \rbrack\!\rbrack (\la )
=f'(\la )g(\la )-f(\la )g'(\la ), \ f,g\in C^\infty(\R),\label{prodd2}
\eea
where $C^\infty (\R )$ denotes the space of smooth functions defined
over $\R$.
It is known that $({\cal B}^q, [\cdot,\cdot])$ and $ (C^\infty (\R), 
\lbrack\!\lbrack \cdot, \cdot \rbrack\!\rbrack) $ are all Lie algebras.

We now assume that $U\in \widetilde {\cal V}^r$ and 
the Gateaux derivative operator $U'$ is injective throughout the paper.
Let us consider the discrete spectral problem
\be \left \{ \ba {l} E\phi =U\phi=U(n,u, \la )\phi ,
 \vspace{2mm} \\ \phi_t=V\phi=V(n,u,\la )\phi, \ea \right.
\label{sp}\ee 
where $V\in {\widetilde {\cal V}}^r.$
Its adjoint system reads as 
\[ \left \{ \ba {l}
E^{-1}\psi =U\psi=U(n,u,\la )\psi,
 \vspace{2mm} \\ \psi_t=(EV)\psi=(EV(n,u,\la ))\psi. \ea \right.
\]
Their integrability conditions 
are given by the following discrete zero curvature
equation 
\be U_t=(EV)U-UV.\label{ic}\ee 
If the operator equation
(\ref{ic}) is equivalent to a system of 
discrete evolution equations $u_t=K(n,u),\ K 
\in {\cal B}^q$, then it is called a discrete
zero curvature representation of $u_t=K(n,u)$.
Evidently 
\[ U_t=U'[u_t]+f(\la )U_\la ,\ 
\textrm{if}\  \la _t=f(\la ),\]
where $U_\la =\frac {\part U}{\part \la }$.
Therefore  a system of 
discrete evolution equations $u_t=K(n,u),\ K\in {\cal B}^q$, 
is the integrability
condition of (\ref{sp}) with the evolution law 
$\la _t=f(\la )$ if and only if 
\be U'[K]+fU_\la =(EV)U-UV.\label{zcressentialrelation}\ee
Note that the injective property of $U'$ is indispensable
in deriving zero curvature representations of systems of evolution equations.
The equation (\ref{zcressentialrelation})
exposes an essential relation between a system of 
discrete evolution equations and 
its discrete zero curvature representation. It 
will play an important role in the context of our construction
of master symmetries.

The paper is divided into five sections. The next section will be devoted
to a general algebraic structure related to 
discrete zero curvature equations.
Then the third section will establish 
an approach for constructing master symmetries by the use of 
discrete zero curvature representations, along with an explanation of
why there exist master symmetries for systems of discrete evolution equations.
In the fourth section, we will go on to illustrate 
our approach by three concrete examples of 
lattice hierarchies. Finally, the fifth section 
provides a conclusion and some remarks. 

\section{Basic algebraic structure}

We aim to discuss Lie algebraic structures of symmetries including
master symmetries
by using zero curvature equations.
It is natural to ask what algebraic structure exists,
related to zero curvature equations.
To answer this question, we first plan to expose a 
Lie algebraic structure
for the space $({\cal B}^q, \widetilde {\cal V}^r, C^\infty (\R))$.

Let $(K,V,f),
\,(S,W,g)\in ({\cal B}^q, \widetilde {\cal V}^r, C^\infty (\R))$,
in other words, $K,S$ are vector fields, $V,W$ are  $r\times r$ matrix 
operators and $f,g$ are smooth functions.
We introduce their product
\be \lbrack\!\lbrack (K,V,f), (S,W,g)\rbrack\!\rbrack =([K,S],
\lbrack\!\lbrack V, W\rbrack\!\rbrack,
\lbrack\!\lbrack f, g\rbrack\!\rbrack),\label{prodds}\ee 
where $[K,S], 
\lbrack\!\lbrack f, g\rbrack\!\rbrack$ are defined by (\ref{prodd1}),
 (\ref{prodd2}), respectively,  and 
$\lbrack\!\lbrack V, W\rbrack\!\rbrack$ is defined by 
\be \lbrack\!\lbrack V, W\rbrack\!\rbrack =V'[S]-W'[K]+[V,W]+gV_\la 
-fW_\la ,\label{prodd3}\ee 
where $ [V,W]=VW-WV$. The same product as (\ref{prodd3}) 
has been introduced for the continuous case in \cite{Ma1993}.

\begin{thm} \label{Liealgebraoftriple}
{\rm (Lie algebra)} The space
$(({\cal  B} ^q, {\widetilde  {\cal  V}}^r, C^\infty (\R )), 
\lbrack\!\lbrack \cdot,\cdot \rbrack\!\rbrack) $ 
 is a Lie algebra,
the product
$\lbrack\!\lbrack \cdot,\cdot \rbrack\!\rbrack$ being
defined by (\ref{prodds}), i.e.,
\[
\lbrack\!\lbrack (K,V,f), (S,W,g)\rbrack\!\rbrack =([K,S],
\lbrack\!\lbrack V, W\rbrack\!\rbrack,
\lbrack\!\lbrack f, g\rbrack\!\rbrack)\]
where \[\left\{\ba {l}
[K,S]=K'[S]-S'[K],
   \vspace{2mm}\\
 \lbrack\!\lbrack V, W\rbrack\!\rbrack =V'[S]-W'[K]+[V,W]+gV_\la
-fW_\la ,\vspace{2mm}\\
  \lbrack\!\lbrack f, g\rbrack\!\rbrack  (\lambda )
  =f'(\lambda )g(\lambda )- f(\lambda )g'(\lambda ) .\ea
\right.\]
\end{thm}

The proof of the theorem will be given in Appendix A. Upon looking at 
the product a little bit more carefully, we can find that
the Lie algebra $(({\cal B}^q, {\widetilde {\cal V}}^r, C^\infty (\R)), 
\lbrack\!\lbrack \cdot, \cdot \rbrack\!\rbrack)$ has a Lie sub-algebra 
$(({\cal B}^q, {\widetilde {\cal V}}^r, 0), 
\lbrack\!\lbrack \cdot, \cdot \rbrack\!\rbrack)$, for which everything
corresponds to the isospectral case.
Moreover the center of an element of this Lie sub-algebra is often Abelian.

The above theorem exposes that a Lie algebraic structure hidden
in the back of vector fields, Lax operators and spectral evolution laws.
Usually we just touch Lie algebraic structures of vector fields
while discussing symmetries. If we analyze symmetries
from the point of zero curvature equations, 
it is natural that we need to find and
handle
Lie algebraic structure for all triples $(K,V,f)
\in ({\cal B}^q, \widetilde {\cal V}^r, C^\infty (\R))$ where $K,V$ and $f$
are related to each other by zero curvature equations.
In other words, we need to observe
how two triples
$(K,V,f), (S,W,g)$ appeared in
zero curvature equations connect with each other.
The following theorem tells us that such a kind of connection
can be reflected by the Lie algebraic operation of
$ ({\cal B}^q, \widetilde {\cal V}^r, C^\infty (\R))$ in Theorem
\ref{Liealgebraoftriple}. Its proof can be found in Appendix B.

\begin{thm} {\rm (Algebraic structure of representations)}
\label{AlgS}
Let $V,W\in {\widetilde  {\cal  V}}^r $, $K,S\in {\cal  B}^q$
 and $f,g\in C^\infty(\R )$.
If two equalities  
\bea && (EV)U-UV=U'[K]+fU_\la ,\label{star1}\\ && (EW)U-UW=U'[S]+gU_\la ,
\label{star2}
\eea
hold, then we have a third equality
\be (E\lbrack\!\lbrack V,W\rbrack\!\rbrack)U-U\lbrack\!\lbrack V,W
\rbrack\!\rbrack=U'[T]+\lbrack\!\lbrack f,g\rbrack\!\rbrack U_\la ,\
{\rm where}\ T=[K,S] ,\ee
where $\lbrack\!\lbrack V,W\rbrack\!\rbrack$, $[K,S]$ and $
 \lbrack\!\lbrack f,g\rbrack\!\rbrack$ are defined by (\ref{prodd3}),
    (\ref{prodd1}) and (\ref{prodd2}), respectively.
\end{thm}

According to this theorem, we can easily find that 
if a system $u_t=K(n,u)$ is isospectral, i.e., $\la _t=f=0$, then 
the product system $u_t=[K,S]$ for any $S\in {\cal B}^q$
can be viewed to be still isospectral
because we have $\lbrack\!\lbrack f,g\rbrack\!\rbrack =
\lbrack\!\lbrack 0,g\rbrack\!\rbrack = 0$, where $g$ is the evolution law
corresponding $u_t=S(n,u)$.
Actually the above theorem gives a discrete zero curvature representation
for a product system $u_t=[K,S]$,
which possesses the same order matrix operators as ones for 
the original systems 
$u_t=K(n,u)$ and $u_t=S(n,u)$
(see \cite{Ma1993} for the continuous case).
Combining two theorems above can show the following.

\begin{cor} The space defined by
\[ \{(K,V,f)\in ({\cal B}^q , {\tilde {\cal V}}^r , C^\infty (\R ))|\, 
U'[K]+fU_\la =(EV)U-UV\} \]
is a Lie sub-algebra of 
$({\cal  B} ^q, {\widetilde  {\cal  V}}^r, C^\infty (\R ))$ 
under the Lie product (\ref{prodds}). 
\end{cor}

This corollary tells us a Lie
algebraic structure about zero curvature equations
and 
will help us to establish Lie algebraic structures
of symmetries including master symmetries. 

However, for zero curvature representations,
some interesting problems remain to be solved. For example,
assuming that two initial
systems $u_t=K(n,u)$ and $u_t=S(n,u)$ have  
zero curvature representations possessing different order matrix operators,
we want to know whether there exist any zero curvature representations
for the product system $u_t=[K,S]$
and what structures the resulting
zero curvature representations possess
if the answer is yes.
It is likely to be helpful in solving this problem 
to use the Kronecker product as \cite{MaG}. 

\section{Lax operators and master symmetries}

Assume that we already have a hierarchy of 
isospectral integrable systems of discrete evolution
equations of the form
\be u_{t}
=K_k=\Phi ^kK_0,\ \Phi \in {\cal  V}^q,\ K_0\in {\cal  B}^q, \ k\ge 0.
\label{dlh1}\ee
or of the form 
\be   u_{t}=K_k=JG_k=MG_{k-1},\ J,M \in {\cal  V}^q,
\ G_{k-1}\in {\cal  B}^q, \ k\ge 0.\label{dlh2}
\ee
associated with a discrete spectral problem 
\be E\phi =U\phi ,\ \phi =(\phi_1,\cdots, 
\phi_r)^T.\label{dsp}\ee 
The second form (\ref{dlh2}) more often occurs 
than the first form (\ref{dlh1}),
although it is simpler to deal with the first form (\ref{dlh1}).
Generally speaking, 
the operator $\Phi$ above is a hereditary symmetry operator 
(see \cite{Fuchssteiner1979} for definition)
determined
by the spectral problem (\ref{dsp}) and $J,M$ constitute a bi-Hamiltonian 
pair \cite{Magri} \cite{GelfandD}. 
If we choose $\Phi =MJ^{-1}$ when $J$ is invertible, 
then the form (\ref{dlh2})
may be changed into the form (\ref{dlh1}). 
Usually $\Phi $ involves nonlocal operators, for example, 
$\Delta ^{-1}$, but $J,M$ often involves only local operators.
Our examples are all local Hamiltonian systems.

\subsection{Structures of Lax operators}
For a given $X\in {\cal  B}^q$ or $G\in {\cal  B}^q$, let us 
introduce an operator equation of $\Omega\in \widetilde {\cal  V}^r $:
\be 
(E\Omega (X))U-U\Omega (X)=U'[\Phi X]-\lambda U'[X],\label{cq1}
  \ee
in the case of (\ref{dlh1}),
or an operator equation of $\Omega _J\in \widetilde {\cal  V}^r $:
\be 
(E\Omega _J(G))U-U\Omega _J(G)=U'[MG]-\lambda U'[JG]\label{cq2},
  \ee
in the case of (\ref{dlh2}).
We call them the characteristic operator equations of $U$.
The introduction of the operator equation (\ref{cq1}) (or (\ref{cq2})) 
is an important step in our manipulation. 
Obviously, we can choose $\Omega _J(G)=\Omega(JG)$ when $\Phi =MJ^{-1}$.
We demand that (\ref{cq1}) (or (\ref{cq2})) has solutions, 
and $\Omega=\Omega(X)$ (or $\Omega _J(G))$ is a
particular solution at $X$ (or at $G$). Usually 
(\ref{cq1}) (or (\ref{cq2})) has infinitely many solutions once one solution
exists, because we can construct others $\Omega (X)+fV$ for any 
$f\in C^{\infty}(\R)$ when
 $V\in {\cal V}^r\otimes
C[\lambda,\lambda ^{-1}]$
solves the stationary discrete zero curvature equation $(EV)U-UV=0$.
The existence of solutions of $(EV)U-UV=0$
may be resulted from the existence of an isospectral hierarchy associated with 
$E\phi =U\phi $.

\begin{thm} {\rm (Structure of Lax operators)} \label{VWE} 
Let two matrices $V_0, W_0\in\widetilde 
{\cal  V}^r$ and two vector fields 
$K_0,\rho_0\in {\cal B}^q$ (or $\rho_0=J\gamma _0, \ 
\gamma _0\in {\cal B}^q$)
satisfy 
\bea &&(EV_0)U-UV_0=U'[K_0],\label{iv0}\\
 && (EW_0)U-UW_0=U'[\rho _0]+\la U_\la .\label{iw0}\eea  
If we define  $\rho_l,\ l\ge1,$ 
$V_k,\, k\ge 1,$ and $W_l,\, l\ge 1,$ as follows
\bea && \rho_l=\Phi ^l\rho_0,\ l\ge1\  (\it {or}\ 
\rho_l=J\gamma _l=M\gamma _{l-1},\ \gamma _{l}\in {\cal B}^q,\ 
l\ge1), \label {nonidlh}\\
&& V_k=
 \lambda ^k V_0 + \sum _{i=1}^k
\lambda ^{k-i}\Omega (K_{i-1})\ (\it {or}\ \Omega _J(G_{i-1}))
,\ k\ge 1, \label{vk}\\ &&W_l=
 \lambda ^l W_0 + \sum _{j=1}^l
\lambda ^{l-j}\Omega (\rho _{j-1})\ (\it {or}\ \Omega _J(\gamma _{j-1})) 
,\  l\ge 1,
\eea
then $V_k,W_l,\, k,l\ge 0$, satisfy
\be (EV_k)U-UV_k=U'[K_k],\ (EW_l)U-UW_l=U'[\rho_l]+\la ^{l+1}U_\la 
,\ k,l\ge 0. \label{bzce}\ee
Therefore for any $k,l\ge0$, the systems of discrete evolution equations $u_t=K_k$
and $u_t=\rho_l$
possess the isospectral ($\lambda _t=0$) and nonisospectral ($\la 
_t=\la ^{l+1}$) discrete zero 
curvature representations
\[ U_t=(EV_k)U-UV_k,\  U_t=(EW_l)U-UW_l,\]  respectively.
\end{thm}

The theorem shows that the Lax operators associated with two hierarchies of 
interesting vector fields can be constructed simply by a unified form.
Its proof is left to Appendix C.
We are successful, thanks to introducing a characteristic
operator equation. The difficulty
is now transferred to seeking a solution to the characteristic 
operator equation.
However this can automatically be solved on basis of the structure 
of Lax operators of isospectral hierarchies, which will be seen in the 
next subsection.

\subsection{A method for constructing master symmetries}
Now we focus our attention on the construction problem of master symmetries.
Theorem \ref{VWE} already 
shows the structure of Lax operators associated with
the isospectral and nonisospectral hierarchies (refer to \cite{Ma1993b} for
the continuous case). When an isospectral 
hierarchy (\ref{dlh1}) (or (\ref{dlh2})) is known, 
the theorem also provides us with a method to 
construct a nonisospectral hierarchy associated
with the discrete spectral problem (\ref{dsp})  
by solving an initial discrete zero 
curvature equation (\ref{iw0}) and solving 
a characteristic operator  equation
 (\ref{cq1}) (or (\ref{cq2})).

However, a solution to  (\ref{cq1}) (or (\ref{cq2}))
may easily be generated by observing the resulting Lax operators.
In fact, we have
\be \Omega (K_k) \ ({\rm {or}}\ \Omega _J(G_k))=V_{k+1}-\la V_k.
\label{omegas}  \ee
This may be checked, say, for the case of (\ref{dlh1}),
 as follows 
\[
 V_{k+1}-\lambda V_k=
\left( \lambda ^{k+1} V_0 + \sum _{i=1}^{k+1}
\lambda ^{k-i+1}\Omega (K_{i-1})\right)-
 \lambda\left(
 \lambda ^k V_0 + \sum _{i=1}^k
\lambda ^{k-i}\Omega (K_{i-1})\right)=\Omega (K_k),
\]
by using (\ref{vk}).
Now by the first equality of (\ref{bzce}),
we may compute the following
\bea  
&& (E\Omega (K_k))U-U\Omega (K_k)
\nonumber \\ &=& 
(EV_{k+1}-\lambda EV_k)U-U(
V_{k+1}-\lambda V_k)
\nonumber \\ &=& 
((EV_{k+1})U-UV_{k+1})- \lambda ((EV_k)U-UV_k)
\nonumber \\ &=& 
U'[K_{k+1}]-\lambda U'[K_k]
=U'[\Phi K_k]-\lambda U'[K_k],
\nonumber
\eea 
for example, for the case of (\ref{dlh1}).
Therefore we see that a possible solution $\Omega (X)$ 
to (\ref{cq1}) (or $\Omega _J(G)$ to (\ref{cq2}))
may be generated by replacing the element $K_k$ (or $G_k$) in 
the equality (\ref{omegas}) with $X$ (or $G$).

The Lax operator 
matrices $V_{k+1}$ and $ V_k$ are known, when the isospectral 
hierarchy has already been found.
Thus we don't have to directly solve the characteristic operator equations
and then the whole process of construction of the nonisospectral
hierarchy becomes {\it an easy task}:
finding $\rho _0,W_0$ to satisfy (\ref{iw0}) and computing $V_{k+1}-\la V_k$
to find a solution to (\ref{cq1}) (or (\ref{cq2})).

The nonisospectral hierarchy (\ref{nonidlh}) is exactly the master 
symmetries that we need to find. The reasons
are that the product systems between the isospectral hierarchy and 
the nonisospectral hierarchy are still isospectral by Theorem \ref{AlgS}
or as we said before in Section II,
and that usually all systems of the isospectral hierarchy 
commute with each other.
Therefore it is because there exists a nonisospectral hierarchy that
there exist master symmetries for isospectral systems of 
discrete evolution equations derived from a given discrete spectral problem.

In the next section, we shall in detail 
illustrate our construction process by three 
concrete examples and establish the 
corresponding centerless Virasoro symmetry algebras.

\section{Applications}

We illustrate only by three examples how to apply 
the method in the last section to construct master 
symmetries
for various lattice hierarchies.  

To make the process clearer, we introduce a conception  
for a given discrete spectral problem $E\phi =U\phi$, which has an injective
Gateaux derivative $U'$.
That is a uniqueness property similar to the one in the continuous
case \cite{MaS}:
\newline
{\centerline {
{\bf if $(EV)U-UV=U'[K]$,  $V\in {\cal V}^r\otimes C[\la ,\la ^{-1}],
\ K\in {\cal B}^q$,
and $V|_{u=0}=0$, then $V=0$}}}
\newline and further $K=0$ by
the injective property of $U'$. 
It means that if an isospectral ($\la _t=0$) Lax operator 
$V$ equals zero at $u=0$, then  so does $V$ itself.
Actually, this property corresponds to the uniqueness of 
an integrable hierarchy associated with a spectral problem $E\phi =U\phi$.
That is to say, when initial conditions 
and constants of inverse difference operators
are fixed (for example, as in (\ref{Deltainverse1}) 
and (\ref{Deltainverse2})), 
the associated isospectral hierarchy is uniquely determined. Most of 
discrete spectral problems share the uniqueness property.
The following three spectral problems are exactly 
examples which share such a property.

\subsection{The Volterra lattice hierarchy}
{\rm 
Let us first consider the following discrete spectral 
problem \cite{ZhangTOF}:
\be E\phi =U\phi, \ U=\left ( \ba {cc} 1&u\vspace {1mm}
\\ \la ^{-1}&0\ea \right ), \ \phi = \left ( \ba {c} \phi _1\vspace{1mm}
\\\phi_2\ea \right ).\label{sp1}\ee
The corresponding isospectral integrable lattice hierarchy reads as
\be 
 u_{t}=K_k=\Phi ^{k}K_0=u(a_{k+1}^{(1)}-a_{k+1}^{(-1)}),
\ K_0=u(u^{(-1)}-u^{(1)}),\ k\ge0. \label{Volterratypelh}\ee
Here the matrix $V=\sum_{i\ge 0}\left ( \ba {cc} a_{i}&uc_{i+1}^{(1)}
\\ c_i &-a_i\ea \right )\la ^{-i}$ solves the stationary discrete zero 
curvature equation $(EV)U-UV=0$, where we choose 
the initial conditions 
\[a_0=\frac12 ,\, c_0=0,\, a_1=-u,\, c_1=1\]
and the hereditary operator $\Phi$ 
is given by
\be
\Phi=u(1+E^{-1})(-u^{(1)}E^2+u)(E-1)^{-1}u^{-1},  \ee
where $(E-1)^{-1}$ is determined by (\ref{inversed}).
It is worth pointing out that each system in (\ref{Volterratypelh})
is local and polynomially dependent on $u$, 
although the hereditary operator $\Phi$ has
nonlocal and non-polynomially dependent features.

The first discrete evolution equation is the Volterra lattice equation
\cite{Volterra}
\[ (u(n))_t=u(n)(u(n-1)-u(n+1)),\]
which is significantly generalized by Bogoyavlensky \cite{Bogoyavlensky}. 
The associated Lax operators are as follows
\be V_{k}= (\la ^{k+1}V)_{\ge 1}+
\left ( \ba {cc} a_{k+1}&0\vspace{2mm}
\\ c_{k+1} & a_{k+1}^{(-1)} \ea \right ), \ k\ge 0,\ee
where $(P)_{\ge 1}$ denotes
the selection of the terms with degrees of $\la $ no less than $1$.
In particular, the initial isospectral Lax operator reads as 
 \be V_0=\left ( \ba {cc} \frac12 \la -u&\la u\vspace{2mm}
\\ 1 &-\frac1 2\la -u^{(-1)}\ea \right ).\ee
The result till here can be obtained from (\ref{sp1}) 
by using a powerful method in \cite{Tu}.
 
We easily obtain the corresponding quantities in the nonisospectral 
($\la _t=\la $) initial discrete zero 
curvature  equation (\ref{iw0}):
\be \rho_0=u,\ W_0=\left ( \ba {cc} \frac12 & 0\vspace{2mm}
\\ 0 &-\frac1 2\ea \right ),
 \ee
and a solution to the characteristic operator  equation (\ref{cq1})
by (\ref{omegas}):
\be \Omega (X)=\left ( \ba {cc} \Omega _{11}(X)&\Omega _{12}(X)
\vspace{2mm}\\
\Omega _{21}(X)&\Omega _{22}(X)\ea \right ),
\ee 
where $\Omega_{ij}(X)$, $i,j=1,2$, are given by 
\bea &&
\Omega _{11}(X)=
(E-1)^{-1}(-u^{(1)}E^2+u)(E-1)^{-1}u^{-1}X\nonumber \\ &&
 \Omega _{12}(X)= \la uE(E-1)^{-1}u^{-1}X\nonumber \\ &&
\Omega _{21}(X)= (E-1)^{-1}u^{-1}X 
\nonumber \\ &&
\Omega _{22}(X)=-\la (E-1)^{-1}u^{-1}X+E^{-1}(E-1)^{-1}
(-u^{(1)}E^2+u)(E-1)^{-1}u^{-1}X\nonumber .
\eea 
Now by Theorem \ref{VWE}, we obtain a hierarchy of nonisospectral discrete evolution
equations $u_t=\rho_l=\Phi^l\rho_0,\ l\ge0$, associated with the spectral
problem (\ref{sp1}).

Let us now consider how to  compute the corresponding symmetry algebra. 
The idea below can be applied to other cases.
We first make the following computation at $u=0$:
\bea && K_k|_{u=0}=0,\ \rho _l|_{u=0}=
\Phi ^l\rho_0|_{u=0}=0
,\ k,l\ge0,\nonumber\\
&&
V_k|_{u=0}=\la ^k\left (\ba {cc} \frac12 \la & 0\vspace{2mm}\\ 1& 
-\frac 12 \la \ea \right),\ k\ge0,\nonumber\\ && 
W_l|_{u=0}=\la ^l\left (\ba {cc} \frac12  & 0 \vspace{2mm} \\ 
0 & -\frac 12  \ea \right )+
(1-\delta_{l0}) \la ^{l-1} 
\left (\ba {cc} 0 & 0 \vspace{2mm}  \\   
{[n]}& -\la [n]  \ea \right )
,\ l\ge0,\nonumber\\ &&
V_{k\la }|_{u=0}=k\la ^{k-1}\left (\ba {cc} \frac12 \la & 0\vspace{2mm}\\ 1& 
-\frac 12 \la \ea \right )
+\la ^k\left (\ba {cc} \frac12  & 0\vspace{2mm}\\ 0& 
-\frac 12  \ea \right ),\ k\ge0,\nonumber\\&&
W_{l\la }|_{u=0}=l\la ^{l-1}\left (\ba {cc} \frac12  & 0\vspace{2mm}\\ 0& 
-\frac 12  \ea \right )+(1-\delta_{l0})\left (\ba {cc} 0 & 0\vspace{2mm}
\\ (l-1)\la ^{l-2}[n]& 
-l\la ^{l-1}[n] \ea \right )
,\ l\ge0, \nonumber
\eea 
where $V_k,\,W_l,\ k,l\ge0$, are given as in Theorem \ref{VWE} and 
$\delta_{l0}$ represents the Kronecker symbol.
While computing $W_l|_{u=0}$, we need to note that
$\Omega(\rho _0)|_{u=0}\ne 0,$ but 
$\Omega(\rho _l)|_{u=0}= 0,\ l\ge1.$ 
The other two examples below have a similar character, too.  
Now we can find by the definition (\ref{prodd3}) of the product of two
Lax operators that
\be \left \{ \ba {l} 
\lbrack\!\lbrack V_k,V_l\rbrack\!\rbrack|_{u=0}=0,\ k,l\ge 0,\vspace{2mm}\\
\lbrack\!\lbrack V_k,W_l\rbrack\!\rbrack|_{u=0}=(k+1)V_{k+l}|_{u=0},
\ k,l\ge 0,\vspace{2mm}
\\
\lbrack\!\lbrack W_k,W_l\rbrack\!\rbrack|_{u=0}=(k-l)W_{k+l}|_{u=0},
\ k,l\ge 0.\ea \right.\label{ilaxa1}\ee
For example, we can compute that
\bea \lbrack\!\lbrack V_k,W_l\rbrack\!\rbrack|_{u=0} & = &
[V_k|_{u=0},  W_l|_{u=0}]+\la ^{l+1}V_{k\la }|_{u=0}\nonumber\\ &=&
\la ^{k+l}\left [ \left ( \ba {cc} \frac 12 \la & 0 \vspace {2mm}
\\ 1 & -\frac 12 \la  
\ea \right ),\  \left ( \ba {cc} \frac 12  & 0\vspace {2mm}
\\ 0 & -\frac 12  
\ea \right ) +(1-\delta_{l0})\la ^{-1}
\left ( \ba {cc} 0 & 0\vspace {2mm}
\\  {[n]} -\la  [n]
\ea \right )
 \right ]
\nonumber\\ & &
+\la ^{l+1}\left ( k\la ^{k-1}
\left ( \ba {cc} \frac 12 \la & 0\vspace {2mm}
\\ 1 & -\frac 12 \la 
\ea \right )+\la ^k\left ( \ba {cc} \frac 12  & 0\vspace {2mm}
\\ 0 & -\frac 12  
\ea \right )
\right )\nonumber\\ &=&
\la ^{k+l}\left ( \ba {cc} 0  & 0\vspace {2mm}
\\ 1 & 0 
\ea \right )+k\la ^{k+l}\left ( \ba {cc}\frac12 \la  & 0\vspace {2mm}
\\ 1 & -\frac12 \la 
\ea \right )+\la ^{k+l+1}\left ( \ba {cc} \frac 12  & 0\vspace {2mm}
\\ 0 & -\frac 12  
\ea \right )\nonumber\\ &=&
(k+1)\la ^{k+l}\left ( \ba {cc}\frac12 \la  & 0\vspace {2mm}
\\ 1 & -\frac12 \la 
\ea \right )=(k+1)V_{k+l}|_{u=0}.\nonumber
\eea
Because $\lbrack\!\lbrack V_k,V_l\rbrack\!\rbrack, 
\lbrack\!\lbrack V_k,W_l\rbrack\!\rbrack-(k+1)V_{k+l}, 
\lbrack\!\lbrack W_k,W_l\rbrack\!\rbrack-(k-l)W_{k+l},\ k,l\ge0,
$ are all isospectral ($\la _t=0$) 
Lax operators belonging to ${\cal V}^2\otimes
 C[\la ,\la ^{-1}]$  by Theorem \ref{AlgS}, based upon (\ref{ilaxa1}) 
we obtain a Lax operator algebra
by the uniqueness property of the spectral problem (\ref{sp1})
\be \left \{ \ba {l}
\lbrack\!\lbrack V_k,V_l\rbrack\!\rbrack =0,\ k,l\ge 0,\vspace{2mm}\\
\lbrack\!\lbrack V_k,W_l\rbrack\!\rbrack =(k+1)V_{k+l},
\ k,l\ge 0,\vspace{2mm}\\
\lbrack\!\lbrack W_k,W_l\rbrack\!\rbrack=(k-l)W_{k+l},
\ k,l\ge 0.
\ea\right.\label{laxa1}\ee
Further, due to the injective property of $U'$,
we finally obtain a vector field algebra of the isospectral hierarchy 
and the nonisospectral hierarchy
\be \left \{ \ba {l}
\left[ K_k,K_l \right] =0,\ k,l\ge 0,\vspace{2mm}\\
\left[ K_k,\rho _l \right] =(k+1)K_{k+l},
\ k,l\ge 0,\vspace{2mm}\\
\left[ \rho _k,\rho _l\right]=(k-l)\rho _{k+l},
\ k,l\ge 0.
\ea \right. \label{sa1} \ee
This implies that $\rho _l,\ l\ge 0$, are all master symmetries of 
each lattice equation $u_t=K_{k_0}$ 
in the isospectral hierarchy, and the symmetries 
\[ K_k,\,k\ge 0,\ {\rm and}\  
\tau _l^{(k_0)}=t[K_{k_0},\rho_l]+\rho_l,\,l\ge0, \]
constitute  a symmetry algebra of Virasoro type possessing
the same commutator relations as (\ref{sa1}).
}

\subsection{The Toda lattice hierarchy}
Let us second consider the discrete spectral problem \cite{Tu}:
\be E\phi =U\phi, \ U=\left ( \ba {cc} 0&1\vspace {1mm}
\\ -v&\la -p\ea \right ), \ u=\left ( \ba {c} p\vspace{1mm}
\\ v\ea \right ),\ 
\phi = \left ( \ba {c} \phi _1\vspace{1mm}
\\\phi_2\ea \right ).\label{sp2}\ee
The corresponding isospectral integrable Toda lattice hierarchy 
\cite{Kupershmidt}
reads as
\be 
 u_{t}=K_k=\Phi ^{k}K_0=\left ( \ba {c} a_{k+2}-a_{k+2}^{(1)}\vspace{1mm}
\\ v(b_{k+2}^{(1)}-b_{k+2})\ea \right )
,\ K_0=\left ( \ba {c} v-v^{(1)}\vspace{1mm}
\\ v(p-p^{(-1)})\ea \right ),
\ k\ge0. \label{Todah}\ee
Here
$V=\sum_{i\ge0}\left ( \ba {cc} a_i&b_i\vspace{1mm}
\\ -vb_i^{(1)}& -a_i\ea \right )\la ^{-i}$ solves $(EV)U-UV=0$, in which 
we choose 
\[a_0=\frac12 ,\ b_0=0,\ a_1=0, \ b_1=-1,\] 
and the hereditary operator $\Phi$ is determined by
\be
\Phi=\left ( \ba {cc} p&(v^{(1)}E^2-v)(E-1)^{-1}v^{-1}\vspace{1mm}
\\ v(E^{-1}+1)& v(pE-p^{(-1)})(E-1)^{-1}v^{-1}\ea \right ).\ee

The first system of discrete evolution equations 
is the Toda lattice \cite{Toda}
\[ \left \{ \ba {l} (p(n))_t=v(n)-v(n+1),\vspace{2mm}\\
(v(n))_t=v(n)(p(n)-p(n-1)),\ea \right.\]
up to a transform of dependent variables. 
The lattice hierarchy above has a local tri-Hamiltonian structure 
\[ u_{t}=K_k=J\frac {\delta H_{k+2}}{\delta u}=M 
\frac {\delta H_{k+1}}{\delta u}
=N\frac {\delta H_k}{\delta u},\ k\ge 0,\]
where the Hamiltonian operators $J,M,N$ and the 
conserved quantities $H_k$ defined by
\bea 
&& J=\left (\ba {cc} 0 & (1-E)v \\ v(E^{-1}-1)& 0 \ea \right), \nonumber \\  
&& M=J\Phi ^\dagger =-\Phi J=
\left (\ba {cc} Ev-vE^{-1} & p(E-1)v \\ v(1-E^{-1})p& v(E-E^{-1})v 
\ea \right), \nonumber \\
&& N=M\Phi ^\dagger =-\Phi M \nonumber \\
&& =\left (\ba {cc} p(vE^{-1}-Ev)+(vE^{-1}-Ev)p &p^2(1-E)v
+(vE^{-1}-Ev)(1+E)v  \\ v(E^{-1}+1)(vE^{-1}-Ev)+v(E^{-1}-1)p^2 &  
2v(E^{-1}p-pE)v
\ea \right), \nonumber \\
&&H_0=p+\frac1{2}{\rm ln}v,\   H_k=-\frac{b_{k+1}}{k},\  k\ge 1, \nonumber
\eea
where $\Phi ^\dagger $ denotes the conjugate operator of $\Phi $. 
Note that this tri-Hamiltonian structure may be 
established through a trace 
identity \cite{Tu}.
The corresponding Lax operators read as 
\be V_{k}=(\la ^{k+1}V)_++
\left ( \ba {cc} b_{k+2}&0\vspace{1mm}
\\ 0&0\ea \right ), \ k\ge0,\ee 
where the subscript $+$ denotes to select the non-negative part.
Hence in particular 
\be V_0=\left ( \ba {cc} \frac12 \la -p^{(-1)}&-1\vspace{2mm}
\\ v &-\frac1 2\la \ea \right ) .\ee

It is easy to 
find the corresponding quantities in the nonisospectral ($\la _t=\la $)
initial discrete zero 
curvature  equation  (\ref{iw0}):
\be \rho_0=\left ( \ba {c} p\vspace{2mm}
\\ 2v\ea \right ),\ W_0=\left ( \ba {cc} [n]-1 & 0\vspace{2mm}
\\ 0 &[n]\ea \right ),
 \ee 
where $[n]$ is the multiplication operator defined by (\ref{multio}),
and a solution to the 
characteristic operator  equation (\ref{cq1}) by (\ref{omegas}):
\be \Omega (X)=\left ( \ba {cc} \Omega _{11}(X)&\Omega _{12}(X)\vspace{2mm}\\
\Omega _{21}(X)&\Omega _{22}(X)\ea \right ), \ X=
\left ( \ba {c} X_1 \vspace{2mm}\\  X_2 \ea \right ), 
\ee 
where $\Omega_{ij}(X)$, $i,j=1,2$, are given by 
\bea &&
\Omega _{11}(X)=
E^{-1}(E-1)^{-1}X_1+(p^{(-1)}-\la )(E-1)^{-1}v^{-1}X_2,\nonumber \\ &&
 \Omega _{12}(X)= (E-1)^{-1}v^{-1}X_2,\nonumber \\ &&
\Omega _{21}(X)= vE(E-1)^{-1}v^{-1}X_2, 
\nonumber \\ &&
\Omega _{22}(X)= (E-1)^{-1}X_1 \nonumber .
\eea 
In this way, we obtain a hierarchy of  
nonisospectral systems of discrete evolution equations 
$\rho_l=\Phi ^l\rho_o,\ l\ge0$, 
associated with the spectral problem  (\ref{sp2}).

In order to construct a vector field algebra, 
we make a similar computation at $u=0$:
\bea && K_k|_{u=0}=0,\ \rho _l|_{u=0}=
\Phi ^l\rho_0|_{u=0}=0,\ k,l\ge0,\nonumber\\
&&
V_k|_{u=0}=\la ^k\left (\ba {cc} \frac12 \la & -1\vspace{2mm}\\ 0& 
-\frac 12 \la \ea \right),\ k\ge0,\nonumber\\&& 
W_l|_{u=0}=\la ^l\left (\ba {cc} [n]-1  & 0\vspace{2mm}\\ 0& 
[n]  \ea \right )+
(1-\delta_{l0})\left (\ba {cc} -2\la [n]  & 2[n]\vspace{2mm}\\ 0& 
0  \ea \right )
,\ l\ge 0,\nonumber\\&&
V_{k\la }|_{u=0}=k\la ^{k-1}\left (\ba {cc} \frac12 \la &-1\vspace{2mm}\\ 0& 
-\frac 12 \la \ea \right )+\la ^k
\left (\ba {cc} \frac12  &0\vspace{2mm}\\ 0& 
-\frac 12  \ea \right )
,\ k\ge 0,\nonumber\\&&
W_{l\la }|_{u=0}=l\la ^{l-1}\left (\ba {cc} [n]-1  & 0\vspace{2mm}\\ 0& 
[n]  \ea \right )+
(1-\delta_{l0})\left (\ba {cc} -2l\la ^{l-1} [n]  & 2(l-1)\la ^{l-2}
[n]\vspace{2mm}\\ 0& 
0  \ea \right )
,\ l\ge 0
.\nonumber
\eea 
Now we can find through the product definition of 
$\lbrack\!\lbrack \cdot,\cdot\rbrack\!\rbrack$ in (\ref{prodd3}) that
\be \left \{\ba {l}
\lbrack\!\lbrack V_k,V_l\rbrack\!\rbrack|_{u=0}=0,\ k,l\ge 0,\vspace{2mm}\\
\lbrack\!\lbrack V_k,W_l\rbrack\!\rbrack|_{u=0}=(k+1)V_{k+l}|_{u=0},
\ k,l\ge 0,\vspace{2mm}\\
\lbrack\!\lbrack W_k,W_l\rbrack\!\rbrack|_{u=0}=(k-l)W_{k+l}|_{u=0},
\ k,l\ge 0.\ea \right.\ee
A similar argument yields 
a Lax operator algebra
by the uniqueness property of the spectral problem (\ref{sp2})
\be \left \{ \ba {l}
\lbrack\!\lbrack V_k,V_l\rbrack\!\rbrack =0,\ k,l\ge 0,\vspace{2mm}\\
\lbrack\!\lbrack V_k,W_l\rbrack\!\rbrack =(k+1)V_{k+l},
\ k,l\ge 0,\vspace{2mm}\\
\lbrack\!\lbrack W_k,W_l\rbrack\!\rbrack=(k-l)W_{k+l},
\ k,l\ge 0.
\ea\right.\label{laxa2}\ee
And then because of the injective property of $U'$,
we obtain a semi-product Lie algebra of the isospectral hierarchy and 
the nonisospectral hierarchy
\be \left \{ \ba {l}
\left[ K_k,K_l\right] =0,\ k,l\ge 0,\vspace{2mm}\\
\left[K_k,\rho _l\right]
=(k+1)K_{k+l},
\ k,l\ge 0,\vspace{2mm}\\
\left[\rho _k,\rho _l\right]=(k-l)\rho _{k+l},
\ k,l\ge 0,
\ea\right.\label{sa2}\ee 
which gives rise to a symmetry algebra of Virasoro type for the isospectral
Toda hierarchy (\ref{Todah}).

\subsection{A sub-KP lattice hierarchy}
{\rm 
Let us finally consider the discrete spectral problem \cite{WuG}:
\be E\phi =U\phi, \ U=\left ( \ba {ccc} 0&1&0\vspace {1mm}
\\ b-\la &a&1\vspace {1mm}\\ c&0&0\ea \right ),\ 
u= \left ( \ba {c} a\vspace{1mm}
\\ b\vspace{1mm}
\\ c \ea \right ),
 \ \phi = \left ( \ba {c} \phi _1\vspace{1mm}
\\ \phi_2\vspace{1mm}
\\ \phi_3 \ea \right ),\label{sp3}\ee
which is equivalent to $(-E^2+b+aE+E^{-1}c)\phi_1=\la \phi_1,$
a sub-KP discrete spectral problem \cite{BlaszakM}. 
The corresponding isospectral integrable lattice hierarchy reads as
\be 
 u_{t}=K_k=JG_k=MG_{k-1},\  k\ge0, \label{lh3}
\ee
where a Hamiltonian pair $J,M$ and $G_{-1},G_{0},G_1$ are defined by
\bea  J&=&\left (\ba {ccc}E-E^{-1}&0&0\vspace{2mm}\\
0&0&(E^{-1}-1)c\vspace{2mm}\\
0&-c(E-1)&0
\ea\right),\nonumber 
\\  M&=&\left (\ba {ccc}Eb-bE^{-1}+a\Delta _+  
\Delta ^{-1}\Delta _-a&EcE-E^{-1}c&-a\Delta _+ 
\Delta ^{-1}\Delta _-c\vspace{2mm}\\
cE-E^{-1}cE^{-1}&E^{-1}ac-acE&-b\Delta _-c\vspace{2mm}\\
c\Delta _+ -\Delta ^{-1}\Delta _-a&-c\Delta _+  b
&c[ \Delta _+ \Delta ^{-1}\Delta  _- -\Delta _- -\Delta _+ ]c
\ea\right), \ \nonumber \\
G_{-1}&=&\left (\ba {c}0\\1\\0\ea\right),\ G_{0}=
\left (\ba {c}c\\b\\a\ea\right),
\ G_{1}=\left (\ba {c}c(Eb+b)\\b^2+ac+E^{-1}ac
\\a(Eb+b)-Ec-E^{-1}c\ea\right) ,\nonumber 
\eea 
where $\Delta _+,\Delta_-$ are the difference operators:
$\Delta _+=E-1,\ \Delta
 _-=1-E^{-1}$.
The first nonlinear system of discrete evolution equations is 
\[ \left \{ 
\ba {l} (a(n))_t=c(n+1)-c(n-1),\vspace{2mm}\\
(b(n))_t=a(n-1)c(n-1)-a(n)c(n),\vspace{2mm}\\
(c(n))_t=c(n)(b(n)-b(n+1)).
\ea 
\right. \]
We easily find the corresponding quantities in (\ref{iv0}) and (\ref{iw0}):
\bea && K_0=\left ( \ba {c} (E-E^{-1})c\vspace{2mm}
\\ (E^{-1}-1)ac\vspace{2mm}
\\ c(1-E)b\ea \right ), \
 V_0=\left ( \ba {ccc} 0&0&1\vspace{2mm}
\\ c&0&0\vspace{2mm}
\\ -E^{-1}ac &E^{-1}c &\la -b\ea \right ), \nonumber \\
&& \rho_0=J\gamma _0=M
\gamma _{-1}=J
\left ( \ba {c}\frac12 \Delta ^{-1}a\vspace{2mm}
\\-\frac32 [n]\vspace{2mm}
\\ - c^{-1}\Delta _-^{-1}b\ea \right )=M
\left ( \ba {c}0\vspace{2mm}
\\0\vspace{2mm}
\\ -([n]+\frac32 )c^{-1}\ea \right )
=\left ( \ba {c}\frac12 a\vspace{2mm}
\\ b\vspace{2mm}
\\ \frac 32 c\ea \right )
,\nonumber \\ &&
W_0=\left ( \ba {ccc} \frac12 [n]& 0&0\vspace{2mm}
\\ 0 &\frac1 2 ([n]+1)&0\vspace{2mm}\\
0& 0& \frac12 ([n]+2)\ea \right ).
\qquad\quad \qquad \nonumber \eea 
We can also obtain 
a solution to the characteristic operator  equation (\ref{cq2})
by (\ref{omegas}):
\be \Omega _J(G)=\left ( \ba {ccc} 
\Omega _{11}(G)&\Omega _{12}(G)&\Omega _{13}(G)\vspace{2mm}\\
\Omega _{21}(G)&\Omega _{22}(G)&\Omega _{23}(G)\vspace{2mm}\\
\Omega _{31}(G)&\Omega _{32}(G)&\Omega _{33}(G)
\ea \right ),     
\  G=\left ( \ba {ccc} G_{(1)}\vspace{2mm}\\  G_{(2)}\vspace{2mm}\\   
G_{(3)}\ea \right ),  
\ee 
where $\Omega_{ij}(G)$, $i,j=1,2,3$, are determined by 
\bea &&
\Omega _{11}(G)=
-(E^2+E)^{-1}(cG_{(3)}+EaG_{(1)})\nonumber \\ &&
 \Omega _{12}(G)= E^{-1}G_{(1)}\nonumber \\ &&
\Omega _{13}(G)= G_{(2)}\nonumber \\ &&
\Omega _{21}(G)= cEG_{(2)}+(b-\la )G_{(1)} 
\nonumber \\ &&
\Omega _{22}(G)=-(E+1)^{-1}(cG_{(3)}+EaG_{(1)})+aG_{(1)}\nonumber \\&&
\Omega _{23}(G)=G_{(1)}, \\&&
\Omega _{31}(G)=E^{-1}cE^{-1}G_{(1)}-E^{-1}acG_{(2)},\nonumber \\&&
\Omega _{32}(G)=E^{-1}cG_{(2)}\nonumber \\&&
\Omega _{33}(G)=-E(E+1)^{-1}(cG_{(3)}+EaG_{(1)})+\Delta _+ aG_{(1)}-(b -\la )
G_{(2)}.\nonumber
\eea 
By Theorem \ref{VWE}, 
we get a hierarchy of nonisospectral systems of discrete evolution equations
$u_t=\rho_l=\Phi ^l\rho_0,\ l\ge 0$, associated with   the spectral  problem
(\ref{sp3}).

In order to generate a vector field algebra of 
the isospectral hierarchy and the nonisospectral hierarchy, 
we need the following quantities,
which may be directly worked out:
\bea && K_k|_{u=0}=0,\ \rho _0|_{u=0}=J\gamma _0|_{u=0}=0,\ 
\rho _l|_{u=0}=J\gamma _l|_{u=0}=M\gamma 
_{l-1}|_{u=0}=0,
\ k\ge0,\,l\ge1,\nonumber\\
&&
V_k|_{u=0}=\la ^k\left (\ba {ccc} 0 & 0&1\vspace{2mm}\\ 0& 0&0\vspace{2mm}\\ 
0&0& \la \ea \right ),\nonumber\\&& 
W_l|_{u=0}=\la ^l\left (\ba {ccc} \frac12  & 0&0\vspace{2mm}\\ 0& 
\frac 12 ([n]+1)&0 \vspace{2mm}\\ 0& 0& \frac12 ([n]+2)\ea \right )+
(1-\delta_{l0})\la ^{l-1}\left (\ba {ccc} 0 & 0&-\frac32[n]\vspace{2mm}\\ 0& 
0&0\vspace{2mm}\\ 0& 0&-\frac32\la [n]\ea \right ),
\nonumber\\&&
V_{k\la }|_{u=0}=k\la ^{k-1}\left (\ba {ccc} 0 & 0&1\vspace{2mm}\\0 & 0&0
\vspace{2mm}\\ 0& 0&\la \ea \right )+\la ^k\left (\ba {ccc} 0  & 0&0
\vspace{2mm}\\ 0& 
0&0 \vspace{2mm}\\ 0& 
0&1 \ea \right ),\nonumber\\&&
W_{l\la }|_{u=0}=l\la ^{l-1}\left (\ba {ccc} \frac12 [n] & 0&0\vspace{2mm}
\\ 0& 
\frac 12 ([n]+1)&0\vspace{2mm}\\ 0&0&\frac12 ([n]+2) \ea \right )+
(1-\delta_{l0})\left (\ba {ccc} 0 & 0&-\frac32(l-1)\la ^{l-2}[n]
\vspace{2mm}\\ 0& 
0&0\vspace{2mm}\\ 0& 0&-\frac32 l\la ^{l-1}[n]\ea \right )
.\nonumber\qquad\qquad
\eea 
Now we easily find according to the product definition of 
$\lbrack\!\lbrack \cdot,\cdot\rbrack\!\rbrack$ that
\[ \left\{ \ba {l}
\lbrack\!\lbrack V_k,V_l\rbrack\!\rbrack|_{u=0}=0,\ k,l\ge 0,\vspace{2mm}\\
\lbrack\!\lbrack V_k,W_l\rbrack\!\rbrack|_{u=0}=(k+1)V_{k+l}|_{u=0},
\ k,l\ge 0,\vspace{2mm}\\
\lbrack\!\lbrack W_k,W_l\rbrack\!\rbrack|_{u=0}=(k-l)W_{k+l}|_{u=0},
\ k,l\ge 0.\ea \right.\]
The same deduction leads to a Lax operator algebra
\be \left \{ \ba {l}
\lbrack\!\lbrack V_k,V_l\rbrack\!\rbrack =0,\ k,l\ge 0,\vspace{2mm}\\
\lbrack\!\lbrack V_k,W_l\rbrack\!\rbrack =(k+1)V_{k+l},
\ k,l\ge 0,\vspace{2mm}\\
\lbrack\!\lbrack W_k,W_l\rbrack\!\rbrack=(k-l)W_{k+l},
\ k,l\ge 0.
\ea\right .\label{laxa3}\ee
and further a vector field  algebra
\be \left \{ \ba {l}
\left [ K_k,K_l\right] =0,\ k,l\ge 0,\vspace{2mm}\\
\left[ K_k,\rho _l\right] =(k+1)K_{k+l},
\ k,l\ge 0,\vspace{2mm}\\
\left[ \rho _k,\rho _l\right]=(k-l)\rho _{k+l},
\ k,l\ge 0.
\ea\right. ,\label{sa3}\ee
which may generate 
a master symmetry algebra possessing the same algebraic structure as 
(\ref{sa3}).
}

\section{Conclusion and remarks} 
\setcounter{equation}{0}

We have established an algebraic structure related to discrete zero curvature
equations and further introduced a simple but systematic 
approach for constructing master symmetries of first degree for
isospectral lattice 
hierarchies associated with discrete spectral problems.
The resulting algebraic structures also leads to 
an explanation of why there exist master symmetries
of first degree. Some complicated calculation 
in our construction is saved by using 
a characteristic operator equation (\ref{cq1}) (or (\ref{cq2}))
and a uniqueness property of discrete spectral problems.
The crucial step is the construction of the corresponding
nonisospectral lattice hierarchies, which can be 
found by solving an initial nonisospectral discrete zero curvature equation.
Three lattice hierarchies are showed as illustrative examples and 
the corresponding master symmetry algebras of centerless Virasoro type
are exhibited. Some of the results in this paper have been
reported at SIDE II, UK \cite{FuchssteinerM}. 

It is worth noting that three examples described in the last section 
possess the same commutator relations  
between their isospectral and nonisospectral vector fields.
In general, we have $[K_k,\rho_l]=(k+\gamma )K_{k+l},\ 
\gamma=\it {const.}$, but the other two equalities of the whole 
Virasoro algebra don't change. This is also  a common phenomenon 
for continuous integrable hierarchies \cite{ChengL-CSB1991} \cite{Ma1992}.
Furthermore we may add a nonisospectral master symmetry with $\la _t=1$
to the whole Virasoro 
symmetry algebra but this often requires additional checking. 
For example, a nonisospectral master symmetry with $\la _t=1$ of the 
sub-KP lattice hierarchy (\ref{lh3}) is $\rho _{-1}=J\gamma _{-1}=(0,1,0)^T$.
On the other hand, similar to the theory in \cite{Ma1992}, 
we may also choose an operator solution $\Omega (X)$ (or $\Omega_J(G)$)
satisfying $\Omega (X)|_{X=0}=0$ (or $\Omega_J(G)|_{G=0}=0$) 
(all three examples in the last section have this property)
and then we only need to compute 
$\lbrack\!\lbrack V_0,W_0  \rbrack\!\rbrack|_{u=0}$ 
so as to give Lax operator algebras at $u=0$ and finally give Lax operator
algebras generally. 

In our discussion, in fact, 
we haven't used the hereditary property of the recursion operator $\Phi$
(or the bi-Hamiltonian property of $J$ and $M$),
while we construct Virasoro 
symmetry algebras for integrable lattice hierarchies,
and thus it can also be applied to 
lattice hierarchies which possess non-hereditary recursion operators. 
The advantage of our scheme is to fully utilize discrete zero curvature 
equations so that the
whole process to generate master symmetries of first degree 
becomes an easy task.
There were also an algorithm implemented in MuPAD \cite{FuchssteinerIW}
and other direct tricks \cite{OevelZFR} \cite{OevelFZ} \cite{ZhangTOF}
\cite{CherdantsevY} to compute master symmetries
of first degree for systems of discrete evolution equations. 
However our theory focuses on seeking an answer to the existence 
and structure problem
of master symmetries of first degree.

We should mention that there exists a large variety of other theories or 
methods to discuss integrable properties of systems of 
nonlinear discrete equations,
which include Hamiltonian theory
\cite{BruschiR} \cite{RagniscoS}, B\"acklund-Darboux transformation  
\cite{BruschiR-PLA1988} \cite{Oevel1}, R-matrix method \cite{BlaszakM}
\cite{MorrisP}, symmetry reduction 
\cite{LeviW} etc. 
Moreover we can consider the time discretization problem \cite{Suris}
and periodic initial and boundary value problems of time discretizations
\cite{PapageorgiouNC}
for symmetry flows of systems of discrete evolution equations.
The resulting difference equations and mappings should be useful in   
discussing the integrability of the underlying systems of discrete 
evolution equations themselves. 
We are also curious about the following natural problem:  
Are there any higher degree 
master symmetries for systems of discrete evolution equations which don't 
depend explicitly on the evolution variable? 
If the answer is yes,
can we establish any relations between those higher degree 
master symmetries and
discrete zero curvature equations as we did for   
the first degree master symmetries?

\newcommand{\eqnappendix}{
   \renewcommand{\theequation}{A.\arabic{equation}}
   \makeatletter
   \csname $addtoreset\endcsname
   \makeatother}
\eqnappendix

\appendix
\section*{Appendix A: Proof of Theorem \ref{Liealgebraoftriple}}
Let $(K_i,V_i,f_i)\in ({\cal B}^q,{\widetilde
{\cal V}}^r, C^\infty (\R )), \ 1\le i\le 3$. Because the bilinearity and 
the skew-symmetry of the product (\ref{prodds}) are self-evident and 
we already know that the products defined by (\ref{prodd1}) and (\ref{prodd2})
are Lie products,
we only need to prove the following Jacobi identity: 
\be \lbrack\!\lbrack \lbrack\!\lbrack V_1,V_2
\rbrack\!\rbrack ,V_3\rbrack\!\rbrack +{\rm cycle}(1,2,3)=0.\label{b3}\ee
Let us first compute by (\ref{prodd3}) that
\bea &&\lbrack\!\lbrack \lbrack\!\lbrack V_1,V_2
\rbrack\!\rbrack ,V_3\rbrack\!\rbrack =(\lbrack\!\lbrack V_1,V_2
\rbrack\!\rbrack ) '[K_3]-V_3'[[K_1,K_2]]+[
\lbrack\!\lbrack V_1,V_2
\rbrack\!\rbrack  ,V_3]+f_3
\lbrack\!\lbrack V_1,V_2\rbrack\!\rbrack _{\la }- 
\lbrack\!\lbrack f_1,f_2\rbrack\!\rbrack V_{3\la }\nonumber \\
&=& (V_1'[K_2])'[K_3]-(V_2'[K_1])'[K_3]
+[V_1,V_2]'[K_3]+f_2(V_{1\la })'[K_3]- 
f_1(V_{2\la })'[K_3]\nonumber \\ 
&& -V_3'[[K_1,K_2]]+[V_1'[K_2],V_3]-[V_2'[K_1],V_3]+[[V_1,V_2],V_3]
+f_2[V_{1\la },V_3]-f_1[V_{2\la },V_3] \nonumber \\ &&
+ f_3(V_{1 }'[K_2])_{\la }  -f_3(V_2'[K_1])_{\la }+f_3[V_1,V_2]_{\la }
+f_{2\la }f_3V_{1\la }+f_2f_3V_{1\la \la }  \nonumber \\
&& -f_{1\la }f_3V_{2\la }-f_1f_3V_{2\la \la }-
\lbrack\!\lbrack f_1,f_2\rbrack\!\rbrack V_{3\la }.\label{3result}
\eea
We need to use the following fundamental equalities
\bea  
&& (V_{\la })'[K]=(V'[K])_{\la },\ V\in {\widetilde {\cal V}}^r,
\ K\in {\cal B}^q,
\nonumber\\
&& 
[V,W]_{\la }=[V_{\la },W]+[V,W_{\la }], \ V,W\in {\widetilde {\cal V}}^r,
\nonumber \\
&& [V,W]'[K]=[V'[K],W]+[V,W'[K]],\ V,W \in {\widetilde {\cal V}}^r,\ 
K\in {\cal B}^q,  \nonumber  \\
&& V'[T]=(V'[K])'[S]-(V'[S])'[K],\ T=[K,S],
 \ V\in {\widetilde {\cal V}}^r,\ K,S \in {\cal B}^q,
\nonumber \eea
which may be shown by  
a direct computation and
the last equality of which is a similar result as in \cite{Ma1992a}.
Now we can go on to compute that
\bea 
{\Delta _a}^{123}&:= & (V_1'[K_2])'[K_3]-(V_2'[K_1])'[K_3]-V_3'[[K_1,K_2]]
\nonumber\\ 
&= & (V_1'[K_2])'[K_3]-(V_2'[K_1])'[K_3]-(V_3'[K_1])'[K_2]+(V_3'[K_2])'[K_1],
\nonumber\\
{\Delta _b} ^{123} &:= &
[V_1,V_2]'[K_3]+[V_1'[K_2],V_3]-[V_2'[K_1],V_3]\nonumber \\
&=& [V_1'[K_3],V_2]
-[V_2'[K_3],V_1]+[V_1'[K_2],V_3]-[V_2'[K_1],V_3],\nonumber \\ 
 {\Delta_c}^{123} & := & f_2(V_{1\la })'[K_3]-f_1(V_{2\la })'[K_3]
+f_3(V_{1 }'[K_2])_{\la }-f_3(V_2'[K_1])_{\la }\nonumber \\
& =&  f_2(V_{1\la })'[K_3]-f_1(V_{2\la })'[K_3]
+f_3(V_{1 \la })'[K_2])-f_3(V_{2\la })'[K_1],\nonumber \\
{\Delta_d} ^{123} 
&:= & f_2[V_{1\la }, V_3]-f_1[V_{2\la },V_3]+f_3[V_1,V_2]_{\la }
\nonumber \\
& = & f_2[V_{1\la }, V_3]-f_1[V_{2\la },V_3]+f_3[V_{1\la },V_2]
-f_3[V_{2\la },V_1],\nonumber \\ 
{\Delta _e}^{123} &:= &
f_{2\la }f_3V_{1\la }+f_2f_{3}V_{1\la \la }-f_{1\la }f_3V_{2\la }
-f_1f_3V_{2\la \la }- \lbrack\!\lbrack f_1,f_2\rbrack\!\rbrack V_{3\la } 
,\nonumber \\
&=& f_{2\la }f_3V_{1\la }+f_2f_{3}V_{1\la \la }-f_{1\la }f_3V_{2\la }
-f_1f_3V_{2\la \la }- f_{1\la }f_2 V_{3\la }+f_1f_{2\la }V_{3\la } 
.\nonumber 
\eea 
A direct check can result in that 
\[ {\Delta _*}^{123}  +{\rm cycle}(1,2,3)  =0,\  {\rm where}\ 
*=a,b,c,d\ {\rm or}\ e.\]
Noting (\ref{3result}), it follows therefore that 
\bea && \lbrack\!\lbrack \lbrack\!\lbrack V_1,V_2\rbrack\!\rbrack,  V_3 
\rbrack\!\rbrack   +{\rm cycle}(1,2,3)\nonumber \\
&=&{\Delta_a}^{123}+
{\Delta _b}^{123}  
+{\Delta _c}^{123}+ {\Delta _d}^{123}+ {\Delta _e}^{123}
+[[V_1,V_2],V_3]+{\rm cycle}(1,2,3)=0,\nonumber
\eea
which is exactly the Jacobi identity (\ref{b3}) and thus completes 
the proof.

\newcommand{\eqnappendixb}{
   \renewcommand{\theequation}{B.\arabic{equation}}
   \makeatletter
   \csname $addtoreset\endcsname
   \makeatother}
\eqnappendixb
\setcounter{equation}{0}

\appendix
\section*{Appendix B: Proof of Theorem \ref{AlgS}}
The proof is an application of the equalities (\ref{star1}) and 
(\ref{star2}) and the third equality  
\be (U'[K])'[S]-(U'[S])'[K]=U'[T],\ T=[K,S],\label{star3}\ee 
which has been mentioned in the proof of the first theorem.
We observe that 
\[ ({\rm Eqn.}\  (\ref{star1}))'[S]-
({\rm Eqn.}\  (\ref{star2}))'[K]  
+g({\rm Eqn.}\  (\ref{star1}))_{\la }-f({\rm Eqn.}
\  (\ref{star2}))_{\la }.\]
The resulting equality reads as
\bea && (U'[K])'[S]-(U'[S])'[K]+
\lbrack\!\lbrack f,g\rbrack\!\rbrack U_\la \nonumber \\ 
&=& (EV'[S])U+(EV)U'[S]-U'[S]V-UV'[S] \nonumber \\     
&& -(EW'[K])U-(EW)U'[K]+U'[K]W+UW'[K] \nonumber \\     
&& +g(EV_{\la })U+g (EV)U_{\la }-gU_{\la }V-g UV_{\la }\nonumber \\     
&& -f(EW_{\la })U-f (EW)U_{\la }+fU_{\la }W+f UW_{\la }.\label{star4}
\eea
On the other hand, we have immediately
\bea 
&& (E\lbrack\!\lbrack V,W \rbrack\!\rbrack )U-
U\lbrack\!\lbrack V,W \rbrack\!\rbrack \nonumber \\
&=& (EV'[S])U-(EW'[K])U+(EV)(EW)U-(EW)(EV)U 
\nonumber \\ &&
+g(EV_{\la })U-f(EW_{\la })U-UV'[S]+UW'[K]
\nonumber \\
&& -UVW+UWV -gUV_{\la }+fUW_{\la }. \label{star5}
\eea
It follows therefore from (\ref{star3}), (\ref{star4}) and (\ref{star5})
that
\bea &&
(E\lbrack\!\lbrack V,W \rbrack\!\rbrack)U-
U\lbrack\!\lbrack V,W \rbrack\!\rbrack -U'[T] -
\lbrack\!\lbrack f,g \rbrack\!\rbrack U_{\la } \nonumber \\
&=&(E\lbrack\!\lbrack V,W \rbrack\!\rbrack)U-
U\lbrack\!\lbrack V,W \rbrack\!\rbrack -(U'[K])'[S]+(U'[S])'[K] -
\lbrack\!\lbrack f,g \rbrack\!\rbrack U_{\la } \nonumber \\
&=& (EV)\{ (EW)U-V'[S]-gU_{\la }\} -(EW)\{ (EV)U-U'[K]-fU_{\la }\}
\nonumber \\
&& -UVW+UWV +gU_{\la }V -fU_{\la }W +U'[S]V -U'[K]W \nonumber \\
&=& (EV)UW-(EW)UV-UVW+UWV+gU_{\la }V-fU_{\la }W+U'[S]V-U'[K]W
\nonumber \\
&=& \{ (EV)U-UV-fU_{\la }-U'[K]\} W-
\{ (EW)U-UW-gU_{\la }-U'[S]\} V\nonumber \\
&=&0,\nonumber 
\eea
which is what we need to prove.

\newcommand{\eqnappendixc}{
   \renewcommand{\theequation}{C.\arabic{equation}}
   \makeatletter
   \csname $addtoreset\endcsname
   \makeatother}
\eqnappendixc

\appendix
\section*{Appendix C: Proof of Theorem \ref{VWE}}
We prove two equalities in (\ref{bzce}). The rest is obvious. 
We compute that
\bea 
&&(EV_k)U-UV_k \nonumber \\     
&=& \la ^k[(EV_0)U-UV_0]+\sum_{i=1}^k\la ^{k-i}[E\Omega (K_{i-1})U- 
U\Omega (K_{i-1})]\nonumber \\     
&=& \la ^k U'[K_0]+\sum_{i=1}^k\la ^{k-i}\{U'[\Phi K_{i-1}]
-\la U'[K_{i-1}]\}\nonumber \\     
&=& \la ^kU'[K_0]+\sum_{i=1}^k\la ^{k-i}\{ U'[K_i]-\la U'[K_{i-1}]\}
\nonumber \\
& = & U'[K_k],\ k\ge 1;\nonumber \vspace{3mm}\\     
& & (EW_l)U-UW_l \nonumber \\     
&=& \la ^l[(EW_0)U-UW_0]+\sum_{j=1}^l\la ^{l-j}[E\Omega (\rho _{j-1})U- 
U\Omega (\rho_{j-1})]\nonumber \\     
&=& \la ^l\{ U'[\rho _0]+\la U_{\la }\} 
+\sum_{j=1}^l\la ^{l-j}\{ U'[\Phi \rho _{j-1}]-\la U'[\rho _{j-1}]\}
\nonumber \\
&=& \la ^l\{ U'[\rho _0]+\la U_{\la }\} 
+\sum_{j=1}^l\la ^{l-j}\{ U'[\rho _j]-\la U'[\rho _{j-1}]\}
\nonumber \\
&= & U'[\rho _l]+\la ^{l+1}U_{\la },\ l\ge 1.\nonumber    
\eea
Note that we have used the characteristic operator equation (\ref{cq1}) but 
the situation in the case of (\ref{cq2}) is conpletely similar.
The proof is therefore finished.

\vskip 5mm
\noindent{\bf Acknowledgments:} 

The authors are indebted to the referee for invaluable comments.
One of the authors (W. X. Ma) would like to thank the 
Alexander von Humboldt Foundation of Germany,
the City University of Hong Kong and the Research Grants Council of Hong Kong
for financial support. 
He is also grateful to J. Leon, W. Oevel and W. Strampp 
for their helpful and stimulating discussions, 
and to R. K. Bullough and P. J. Caudrey for their warm 
hospitality during his visit at UMIST, UK.


\small
\begin{thebibliography}{99}
\bibitem{MikhailovSS} 
A. V. Mikhailov, A. B. Shabat and V. V. Sokolov, 
in: What is Integrability?, ed. V. E. Zakhalov
(Springer Verlag, Berlin, 1991) p115--184.
\bibitem{LeviY} D. Levi and R. I. Yamilov, 
J. Math. Phys. 38 (1997) 6648--6674.
\bibitem{FuchssteinerF} B. Fuchssteiner and A. S. Fokas,
Physica D 4 (1981) 47--66.
\bibitem{Fokas} A. S. Fokas, 
Studies in Appl. Math. 77 (1987) 253--299.
\bibitem{Fuchssteiner1983} B. Fuchssteiner,
Prog. Theor. Phys. 70 (1983) 1508--1522.  
\bibitem{Ma1990} W. X. Ma, J. Phys. A: Math. Gen.  
23 (1990) 2707--2716.
\bibitem{ChenL} H. H. Chen and Y. C. Lee,
in: Advances in Nonlinear Waves Vol. II, ed. L. Debnath, Research 
Notes in Mathematics 111 (Pitman, Boston, 1985) p233-239.
\bibitem{OrlovS-LMP1986}A. Y. Orlov and E. I. Schulman,
  Lett. Math. Phys. 12 (1986) 171--179.
\bibitem{LiZ-JPA1986}Y. S. Li and G. C. Zhu,
 {\it J. Phys. A: Math. Gen.} 19 (1986) 3713--3725. 
\bibitem{Ma1993c} W. X. Ma, Phys. Lett. A 179 (1993) 179--185; 
Proceedings of the 21st International Conference on the
      Differential Geometry Methods in Theoretical Physics, ed M. L. Ge
(World Scientific, Singapore, 1993) p535-538; 
J. Phys. A: Math. Gen. 25 (1992) L719--L726.
\bibitem{CalogeroD-LNC1978}F. Calogero and A. Degasperis,
Lett. Nuovo Cimento 22 (1978) 420--424.
\bibitem{ChanZ-LMP1987}W. L. Chan and Y. K. Zheng,
  Lett. Math. Phys. 14 (1987) 293--301.
\bibitem{OevelZFR} W. Oevel, H. W. Zhang, B. Fuchssteiner and O. Ragnisco,
J. Math. Phys. 30 (1989) 2664--2670.
\bibitem{OevelFZ} W. Oevel, B. Fuchssteiner and H. W. Zhang, 
Prog. Theor. Phys. 81 (1989) 294--308.
\bibitem{ZhangTOF} H. W. Zhang, G. Z. Tu, W. Oevel and B. Fuchssteiner, 
J. Math. Phys. 32 (1991) 1908--1918.
\bibitem{AblowitzL-JMP19756}
M. J. Ablowitz and J. F. Ladik, 
J. Math. Phys. 16 (1975) 598--603;
J. Math. Phys. 17 (1976) 
1011--1018.
\bibitem{DeviBR-NCA1980}
D. Levi, M. Bruschi and O. Ragnisco,
Nuovo Cimento A,
58 (1980) 56--66.
\bibitem{BruschiLR-JMP1981}
M. Bruschi, D. Levi and O. Ragnisco,
J. Math. Phys. 22 (1981) 2463--2471.
\bibitem{BruschiR-LNC1981}M.
Bruschi and O. Ragnisco,
Lett. Nuovo Cimento 31 (1981) 492--496.
\bibitem{Ma1993} W. X. Ma,  J. Phys. A: Math. Gen.
26 (1993) 2573--2582.
\bibitem{Ma1992a} W. X. Ma, J. Phys. A: Math. Gen. 25 
(1992) 5329--5343.
\bibitem{MaG} W. X. Ma and F. K. Guo, 
Intern. J. Theoret. Phys. 36 (1997) 697--704.
\bibitem{Fuchssteiner1979} B. Fuchssteiner, Nonlinear Analysis TMA 3 (1979)
849--862.
\bibitem{Magri} F. Magri, J. Math. Phys. 19 (1978) 1156--1162.
\bibitem{GelfandD} I. M. Gel'fand and I. Y. Dorfman,
Funct. Anal. Appl. {\bf 13} (1979) 248--262. 
\bibitem{Ma1993b} W. X. Ma, Chinese Science Bulletin 38 (1993)
2025--2031.
\bibitem{MaS} W. X. Ma and W. Strampp, Phys. Lett. A 
185 (1994) 277--286.
\bibitem{Volterra} V. Volterra, Le\c{c}ons sur la Th\'eorie 
Math\'ematique de la Lutte pour la Vie (Gauthier-Villars, Paris, 1931).  
\bibitem{Bogoyavlensky} O. I. Bogoyavlensky, Phys. Lett. A 
134 (1988) 34--38;
USSR Math. Izv. 31 (1988) 435--454.
\bibitem{Tu} G. Z. Tu,  J. Phys. A: Math. Gen. 23 (1990) 3903--3922. 
\bibitem{Kupershmidt} B. G. Kupershmidt, Discrete Lax Equations and
Differential-difference Calculus, Asterisque 123 (1985).
\bibitem{Toda} M. Toda,  Theory of Nonlinear Lattices, 2nd enl. ed. 
(Springer-Verlag, Berlin, 1989).
 \bibitem{WuG} Y. T. Wu and 
X. G. Geng, J. Math. Phys. 37 (1996) 2338--2345.
\bibitem{BlaszakM} M. Blaszak and K. Marciniak,  J. Math. Phys.
 35 (1994)  4661--4682.
\bibitem{FuchssteinerM} B. Fuchssteiner and W. X. Ma,
An approach to master symmetries of lattice equations, to appear in
Proceedings of SIDE II, Canterbury, UK, 1996.
\bibitem{ChengL-CSB1991}
Y. Cheng and Y. S. Li, 
Chinese Sci. Bull. 36 (1991) 
1428--1433.
\bibitem{Ma1992} W. X. Ma,  
J. Math. Phys. 33 (1992) 2464--2476.
\bibitem{FuchssteinerIW} B. Fuchssteiner, S. Ivanov and W. Wiwianka,
Mathematical and Computer Modelling 25 (1997) 91--100.
\bibitem{CherdantsevY} I. Y. Cherdantsev 
 and R. I. Yamilov, Physica D  87 (1995)
 140--144.
\bibitem{BruschiR}M. Bruschi and O. Ragnisco,
J. Math. Phys. 24 (1983) 1414--1421.
\bibitem{RagniscoS} O. Ragnisco and P. M. Santini, Inverse Problems
6 (1990) 441--452;
M Bruschi, O. Ragnisco, P. M. Santini and
G. Z. Tu, Physica D 49 (1991) 273--294.
\bibitem{BruschiR-PLA1988}M. Bruschi and O. Ragnisco,
Phys. Lett. A 129 (1988) 21--25.
\bibitem{Oevel1} W. Oevel, 
in:  Nonlinear Physics: Theory and Experiment, eds. by E. Alfinito,
M. Boiti, L. Martina and F. Pempinelli (World Scientific,
Singapore, 1996) P223--240.
\bibitem{MorrisP} C. Morosi and L. Pizzocchero, J. Math. Phys.  37
(1996) 4484--4513. 
\bibitem{LeviW} D. Levi and P. Winternitz, 
Phys. Lett. A  152 (1991) 335--338;
J. Math. Phys. 34 (1993) 3713--3730.
\bibitem{Suris} Yu. B. Suris, 
J. Phys. A: Math. Gen.  29 (1996) 451--46; J. Math. Phys. 
37 (1996) 3982--3996.
\bibitem{PapageorgiouNC} V. G. Papageorgiou, F. W. Nijhoff and H. W. Capel,
Phys. Lett. A 147 (1990) 106--114;
H. W. Capel, F. W. Nijhoff and V. G. Papageorgiou,  
Phys. Lett. A 155 (1991) 377--387.

\end{thebibliography}
\end{document}